\documentclass[11pt]{article}
\usepackage{graphicx}
\usepackage{cite}
\usepackage{epstopdf}

\begin{document}

\title{\bf Single-spin asymmetries in inclusive deep inelastic \\
scattering and multiparton correlations in the nucleon} 

\author{A.~Metz$^{1}$, D.~Pitonyak$^{1}$, A.~Sch\"afer$^{2}$,
 M.~Schlegel$^{3}$, W.~Vogelsang$^{3}$, J.~Zhou$^{2}$
 \\[0.3cm]
{\normalsize\it $^1$Department of Physics, Barton Hall,
  Temple University, Philadelphia, PA 19122, USA} \\[0.15cm]
{\normalsize\it $^2$Institute for Theoretical Physics, Regensburg University,} \\ 
{\normalsize\it Universit\"atsstra{\ss}e 31, D-93053 Regensburg, Germany} \\[0.15cm]
{\normalsize\it $^3$Institute for Theoretical Physics, T\"ubingen University,} \\ 
{\normalsize\it Auf der Morgenstelle 14, D-72076 T\"ubingen, Germany}} 

\date{\today}
\maketitle

\begin{abstract}
\noindent
Transverse single-spin asymmetries in inclusive deep inelastic lepton-nucleon scattering can be generated through multiphoton exchange between the leptonic and the hadronic part of the process.
Here we consider the two-photon exchange and mainly focus on the transverse target spin asymmetry.
In particular, we investigate the case where two photons couple to different quarks.
Such a contribution involves a quark-photon-quark correlator in the nucleon, which has a (model-dependent) relation to the Efremov-Teryaev-Qiu-Sterman quark-gluon-quark correlator $T_F$.
Using different parametrizations for $T_F$ we compute the transverse target spin asymmetries for both a proton and a neutron target and compare the results to recent experimental data.
In addition, potential implications for our general understanding of single-spin asymmetries in hard scattering processes are discussed.
\end{abstract}

%
%
\section{Introduction}
\label{s:intro}
In order to describe deep inelastic lepton-nucleon scattering (DIS), $\ell(k) + N(P) \to \ell(k') + X$, one normally considers the exchange of only one photon between the leptonic and the hadronic part of the process\footnote{Throughout this work we neglect contributions from the Z-boson exchange.}.  
In this one-photon exchange approximation, the cross section for DIS can be expressed in terms of four structure functions: two for unpolarized scattering and two for double-polarized scattering where both beam and target are polarized.
On the other hand, because of parity and time reversal invariance, single-spin observables are forbidden in the one-photon exchange approximation~\cite{Christ:1966zz}.
However, this restriction does not apply if a multiphoton exchange is taken into account.
In this case there can be a nonzero single-spin effect due to a correlation of the type 
\begin{equation}
\varepsilon^{S P k k'} \equiv
\varepsilon_{\mu \nu \rho \sigma} S^{\mu} P^{\nu} k^{\rho} k'^{\sigma} \,, 
\label{e:ssa_corr}
\end{equation}
where $\varepsilon^{\mu\nu\rho\sigma}$ is the totally antisymmetric Levi-Civita tensor. 
The four-vector $S$ may represent the spin-vector of the nucleon, the incoming lepton, or the outgoing lepton. 
The correlation in Eq.~(\ref{e:ssa_corr}) is nonzero provided that $S$ has a component which is normal to the reaction plane.
Therefore, one can have a transverse single-spin asymmetry (SSA) defined through
\begin{equation}
A_{UT} = \frac{d\sigma^{\uparrow} - d\sigma^{\downarrow}}
              {d\sigma^{\uparrow} + d\sigma^{\downarrow}}
       = \frac{d\sigma^{\uparrow} - d\sigma^{\downarrow}}
              {2 \, d\sigma_{unp}} \,,
\label{e:ssa_def}              
\end{equation}
where the numerator is given by the difference of the cross sections when the nucleon's (transverse) spin vector is flipped.
(Note that the corresponding transverse SSA in the case of, e.g., $p^{\uparrow} p \to \pi X$ is often denoted by $A_N$.)
A precise definition of our sign conventions for $A_{UT}$ is given below.
The asymmetry $A_{UT}$ can be expected to be small for it is proportional to the electromagnetic fine structure constant $\alpha_{em} \approx 1/137$.

While for elastic lepton-nucleon scattering multiphoton exchange has already been extensively studied --- see~\cite{Jones:1999rz,Arrington:2011dn} and references therein, our knowledge about multiphoton exchange in DIS is still rather limited, in particular when it comes to the spin asymmetry $A_{UT}$.
Early measurements of $A_{UT}^{p}$ (asymmetry for a polarized proton) were performed at the Cambridge Electron Accelerator~\cite{Chen:1968mm} and at the Stanford Linear Accelerator~\cite{Rock:1970sj}.
These experiments were carried out in the resonance region, and in either case the asymmetry was found to be zero within the error bars.
The recent measurement of $A_{UT}^{p}$ by the HERMES Collaboration~\cite{:2009wj} constitutes the first such study in the DIS region. 
Both an electron and a positron beam were used, and again no evidence for a nonzero effect was found~\cite{:2009wj}.
Moreover, by using a polarized $^3$He target, preliminary data on $A_{UT}^{n}$ (asymmetry for a polarized neutron) were obtained in the E07-013 experiment in Hall A at Jefferson Lab~\cite{E07-013,Katich:2011}, providing for the first time a nonvanishing transverse SSA in DIS on the (few) percent level. 
The challenging task is to describe both the magnitude and the sign of the proton and neutron data.

On the theoretical side there exists an early phenomenological calculation of $A_{UT}^{p}$ which concentrates on the nucleon-pion final state, in a kinematical region where the reaction is dominated by the excitation and the decay of the $\Delta(1232)$ resonance~\cite{Cahn:1970gp}.
More recently an attempt was made to describe transverse SSAs in inclusive DIS in the parton model~\cite{Metz:2006pe}.
That work considered the coupling of the exchanged photons to the {\it same} quark inside the nucleon.
While a compact and well-behaved result for the beam spin asymmetry was obtained, the transverse target SSA turned out to be infrared (IR) divergent~\cite{Metz:2006pe}.
We review these results below in Sec.~\ref{s:same} and explain how the IR divergence can be removed.
If one keeps quark mass effects, the target SSA also receives a contribution involving the transversity distribution of the nucleon~\cite{Afanasev:2007ii}.
An estimate of this effect gave rise to quite small results for both $A_{UT}^{p}$ and $A_{UT}^{n}$~\cite{Afanasev:2007ii}.
We also point out that in Ref.~\cite{Schlegel:2009pw} the influence of two-photon exchange on observables in semi-inclusive DIS has been explored for the first time.

In the present work we mainly focus on the situation when the exchanged photons couple to {\it different} quarks inside the nucleon.
As we are going to argue below, numerically this contribution presumably dominates over the one where the photons couple to the same quark.
The required collinear twist-3 calculation is very similar to the treatment of transverse SSAs in hadron-hadron scattering~\cite{Efremov:1981sh,Qiu:1991pp,Qiu:1998ia,Eguchi:2006qz,Kouvaris:2006zy}.
The analytical result for the target asymmetry depends on a quark-photon-quark correlator ($q \gamma q$-correlator).
By using a valence quark picture of the nucleon, we relate this object to the so-called Efremov-Teryaev-Qiu-Sterman (ETQS) quark-gluon-quark correlator ($qgq$-correlator) $T_F$~\cite{Efremov:1981sh,Qiu:1991pp,Qiu:1998ia}, which plays an important role in the quantum chromodynamics (QCD) description of SSAs.
On the basis of different parametrizations for $T_F$, we then compute $A_{UT}^{p}$ and $A_{UT}^{n}$.
Depending on the input for $T_F$ this approach does provide a reasonable description of the available data.
Our finding also has potential implications for the general understanding of single-spin asymmetries in hard scattering processes.

The paper is organized as follows: in Sec.~\ref{s:same} we consider the two-photon exchange where both photons couple to the same quark.
In that part we briefly review what is already known from the literature~\cite{Metz:2006pe,Afanasev:2007ii}, but also provide some new insights. 
Section~\ref{s:different} then deals with the coupling of the two photons to different quarks.
There we provide the analytical result for the transverse target SSA, the aforementioned relation between the $q \gamma q$ correlator and the ETQS matrix element $T_F$, numerical results for $A_{UT}^{p}$ and $A_{UT}^n$, and some discussion on the implications of our results.
We summarize our work in Sec.~\ref{s:summary}.
\begin{figure}[t]
\begin{center}
\includegraphics[width=14.0cm]{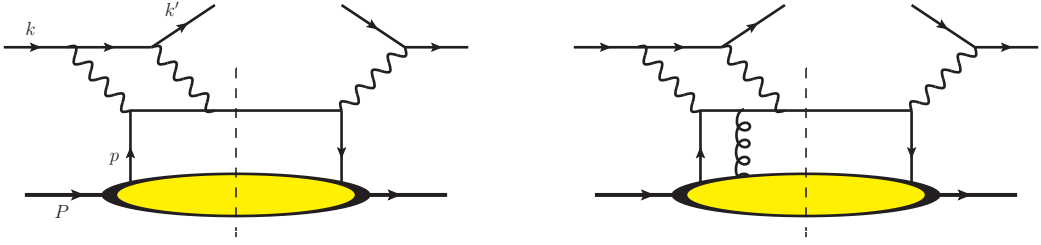} 
\end{center}
\caption{Left panel: Two-photon exchange contribution (box graph) to inclusive DIS in the parton model. 
The Hermitian conjugate diagram, not shown in the figure, has to be considered as well. 
The so-called crossed box graph does not contribute to $A_{UT}$.
Right panel: Sample diagram for two-photon exchange contribution involving a $qgq$ correlator. 
Such diagrams contribute to the leading power of the SSA for a transversely polarized target.} 
\label{f:twogamma}
\end{figure}

%
%
\section{Photons coupling to the same quark}
\label{s:same}
In order to give a reference point and to fix our notation, let us first recall the well-known unpolarized cross section for inclusive DIS in the parton model in the one-photon exchange approximation,
\begin{equation} 
k^{\prime 0} \frac{d\sigma_{unp}}{d^3\vec{k}'} =  
\frac{4 \alpha_{em}^2}{Q^4 \, y} \, \bigg( 1 - y + \frac{y^2}{2} \bigg) 
\sum_q e_q^2 \, x f_1^q(x) =
\frac{2 \alpha_{em}^2 \, y}{Q^4} \, \frac{\hat{s}^2 + \hat{t}^2}{\hat{u}^2} 
\sum_q e_q^2 \, x f_1^q(x) \,,
\label{e:unp}
\end{equation}
where $f_1^q$ denotes the unpolarized twist-2 quark distribution for a quark flavor $q$.
A summation over both quarks and antiquarks is understood in Eq.~(\ref{e:unp}).
We make use of the common DIS variables
\begin{equation}
Q^2 = - (k - k')^2 \,, \qquad
x = \frac{Q^2}{2 P \cdot (k - k')} \,, \qquad
y = \frac{P \cdot (k - k')}{P \cdot k} \,.
\label{e:dis_var}
\end{equation}
Upon neglecting the nucleon mass, the variables in~(\ref{e:dis_var}) are related through $y = Q^2 / (xs)$, with $s = 2 P \cdot k$ denoting the squared center-of-mass energy of the reaction.
The (longitudinal) momentum of the struck quark is given by $p = x P$ (see also Fig.$\,$\ref{f:twogamma}).
In Eq.~(\ref{e:unp}) we also use the partonic Mandelstam variables for the elastic lepton-quark scattering,
\begin{equation}
\hat{s} = (xP + k)^2 = \frac{Q^2}{y}\,, \qquad
\hat{t} = (xP - k')^2 = - \frac{Q^2 (1-y)}{y}\,, \qquad
\hat{u} = (k - k')^2 = - Q^2 \,,
\end{equation}
which satisfy $\hat{s} + \hat{t} + \hat{u} = 0$.

We now turn to the transverse SSA for a polarized lepton beam. 
A nonzero spin asymmetry arises when taking into account the two-photon exchange contribution shown on the left panel of Fig.$\,$\ref{f:twogamma}.
The essential element of the calculation is the imaginary part of the lepton-quark box diagram which, in principle, was already computed in 1960~\cite{Barut:1960zz}.
The real part of that diagram does not contribute to the SSA.
The result for the spin-dependent cross section reads~\cite{Metz:2006pe}
\begin{equation} 
k^{\prime 0} \frac{d\sigma_{pol}^{\ell}}{d^3\vec{k}'} =  
\frac{4 \alpha_{em}^3}{Q^8} \, m_{\ell} \, x y^2 \, \varepsilon^{S_{\ell} P k k'}
\, \sum_q e_q^3 \, x f_1^q(x) \,,
\label{e:pol_lepton}
\end{equation}
where we use the shorthand notation of Eq.~(\ref{e:ssa_corr}) and the convention $\varepsilon^{0123} = 1$.
The expression in~(\ref{e:pol_lepton}) is proportional to the lepton mass $m_{\ell}$, and, therefore, the general behavior of the spin asymmetry is given by $A_{UT}^{\ell} \sim \alpha_{em} m_{\ell} / Q$.
This implies that even for DIS with a muon beam $A_{UT}^{\ell}$ should at best be of $\mathcal{O}(10^{-3})$.
Although the asymmetry is suppressed like $1/Q$, the leading twist unpolarized quark distributions for the nucleon enter, while the suppression is caused by the lepton side of the process.

Computing $A_{UT}$ for a nucleon target on the basis of the diagram on the left panel of Fig.$\,$\ref{f:twogamma} is more involved since this observable is a genuine twist-3 effect.
One has contributions related to (i) collinear twist-3 correlators in the nucleon, (ii) transverse quark motion, and (iii) the quark mass.  
The calculation provides the cross section~\cite{Metz:2006pe}\footnote{The contribution depending on the quark mass $m_q$ in Eq.~(\ref{e:pol_nucleon_1}) was not yet given in~\cite{Metz:2006pe}.}
\begin{eqnarray} 
k^{\prime 0} \, \frac{d\sigma_{pol}^{N}}{d^3\vec{k}'} & = & 
\frac{4 \alpha_{em}^3}{Q^8} \, \frac{M x^2 y}{1-y} \, \varepsilon^{S_{N} P k k'} 
\sum_q e_q^3 
\bigg[ \bigg( x g_T^q(x) - g_{1T}^{(1) q}(x) - \frac{m_q}{M} h_1^q(x) \bigg)
\nonumber \\
& & \mbox{} \times 
\bigg( (1-y)^2 \ln \frac{Q^2}{\lambda^2} + y(2-y) \ln y + y(1-y) \bigg) 
+ \frac{m_q}{M} h_1^q(x) y (1-y) \bigg] + \ldots \,,
\label{e:pol_nucleon_1}
\end{eqnarray}
which is proportional to the nucleon mass $M$.
This result contains the collinear twist-3 two-parton correlator $g_T$ which can also be measured, for instance, through the longitudinal-transverse double spin asymmetry $A_{LT}$ in inclusive DIS.
The contribution due to transverse quark motion is described by the correlator $g_{1T}^{(1)}$.
This correlation function represents a particular moment of the transverse momentum dependent parton distribution $g_{1T}$~\cite{Mulders:1995dh,Bacchetta:2006tn},
\begin{equation}
g_{1T}^{(1)}(x) = \int d^2\vec{p}_T \, \frac{\vec{p}_T^{\;2}}{2 M^2} \, g_{1T}(x,\vec{p}_T^{\;2}) \,. 
\end{equation}
In Eq.~(\ref{e:pol_nucleon_1}) the term proportional to the quark mass is described by the transversity distribution $h_1$~\cite{Ralston:1979ys}.
We point out that the transversity contribution to $d\sigma_{pol}^{N}$ was first published in Ref.~\cite{Afanasev:2007ii}.
In that work a projection operator for transversity was used which contains $m_q$.
Then the calculation becomes identical to the case of a polarized lepton beam discussed above, and the full transversity-related result is just given by the very last term in Eq.~(\ref{e:pol_nucleon_1}).
In contrast, the transversity contribution in~(\ref{e:pol_nucleon_1}) is obtained using a projection operator without a quark mass term.
As we argue below, the {\it complete} result for the spin-dependent cross section does not depend on the choice of this projector. 

The calculation leading to~(\ref{e:pol_nucleon_1}) satisfies electromagnetic gauge invariance, yet the result contains an uncancelled IR divergence.
This divergence can be regularized by a photon mass $\lambda$ which shows up in the logarithm $\ln (Q^2/\lambda^2)$~\cite{Metz:2006pe}.
(Terms vanishing in the limit $\lambda \to 0$ are not listed in~(\ref{e:pol_nucleon_1}).)
This feature clearly hints at additional contributions that also have to be taken into account.
Indeed, since we are dealing with a twist-3 observable, one has to consider $qgq$ correlations in the nucleon.
The dots in~(\ref{e:pol_nucleon_1}) indicate these missing contributions, and a sample diagram is shown on the right panel of Fig.$\,$\ref{f:twogamma}. (Diagrams where one of the exchanged photons is {\it real} can also contribute to the target SSA. However, they do not matter for the mentioned IR divergence.)
It has been speculated that the inclusion of such terms could ultimately generate an IR-finite result~\cite{Metz:2006pe,Afanasev:2007ii}.
The twist-3 correlator which is relevant for the discussion of the IR divergence is denoted by $\tilde{g}_T$ in Refs.~\cite{Mulders:1995dh,Bacchetta:2006tn}.
Through QCD equations of motion this function is related to the parton correlators showing up in~(\ref{e:pol_nucleon_1})~\cite{Mulders:1995dh,Bacchetta:2006tn},
\begin{equation}
x \tilde{g}_T(x) = x g_T(x) - g_{1T}^{(1)}(x) - \frac{m_q}{M} h_1(x) \,.
\label{e:eom}
\end{equation}
Since the IR-divergent term in Eq.~(\ref{e:pol_nucleon_1}) appears with exactly the linear combination of parton correlators showing up on the r.h.s.~of Eq.~(\ref{e:eom}), it is justified to expect that the inclusion of the $\tilde{g}_T$ term will provide an IR-finite result.
In fact, we were able to show that the IR divergence indeed cancels after taking into account all contributions (with virtual photon exchange), and currently we are working towards a complete result~\cite{Schlegel:prog}.
Let us also mention that the transversity-related quark mass term  can be expected to be small because of the relative prefactor $m_q/M$.
Actually, it was found that both $A_{UT}^{p}$ and $A_{UT}^{n}$, based on the transversity contribution only, are just of $\mathcal{O}(10^{-4})$ even when using a constituent quark mass~\cite{Afanasev:2007ii}.
We note in passing that the quark mass term in~(\ref{e:eom}) is absent if one works with the projection operator for transversity used in~\cite{Afanasev:2007ii}.
This shows that indeed both transversity projectors lead to the same final result for the spin-dependent cross section.
\begin{figure}[t]
\begin{center}
\includegraphics[width=13.0cm]{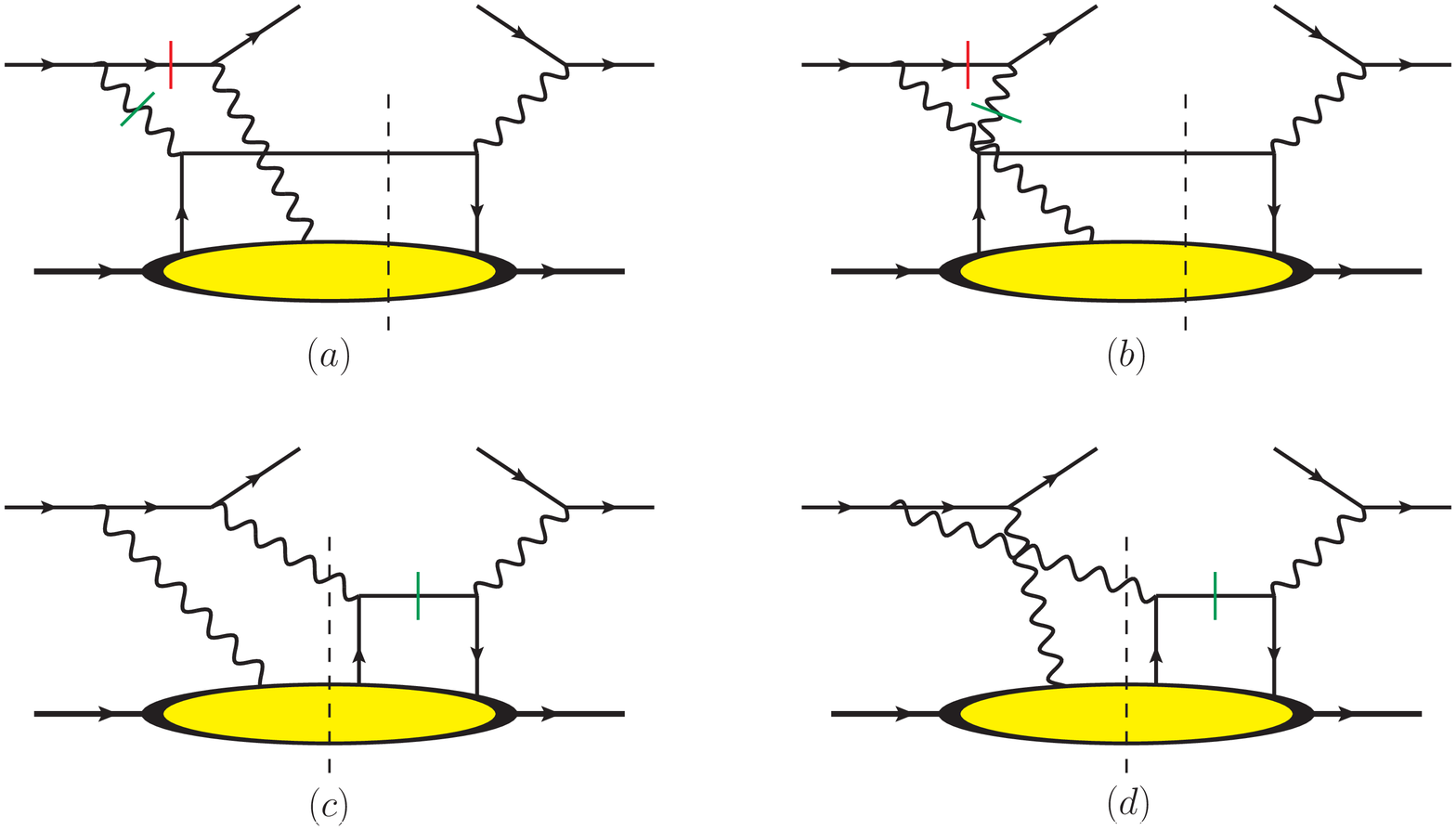} 
\end{center}
\caption{Two-photon exchange contributions to inclusive DIS where the photons couple to different quarks.
(The Hermitian conjugate diagrams are not shown.)
Such contributions can be expressed through a $q \gamma q$ correlator in the nucleon.
Particle lines that can go on-shell are indicated by a short dash (see text for more details).}
\label{f:twogamma_3parton}
\end{figure}

%
%
\section{Photons coupling to different quarks}
\label{s:different}
\subsection{Analytical results}
\label{s:different_1}
Now we consider the case where the two photons couple to different quarks in the nucleon --- see Fig.$\,$\ref{f:twogamma_3parton} for the relevant Feynman diagrams.
For the target SSA such diagrams contribute to the same order in $1/Q$ as those discussed in the previous section. 
Note, however, that they are not relevant for the dominant term of the lepton SSA.
Therefore, the result in Eq.~(\ref{e:pol_lepton}) represents the complete leading power expression for $d\sigma_{pol}^{\ell}$.

Imaginary parts are a necessary condition for the existence of SSAs.
In Feynman graphs they generally arise when internal particles go on-shell, i.e., when one hits the pole of a particle propagator.
For the diagrams in Fig.$\,$\ref{f:twogamma_3parton} there are two sources of such poles: first, the lepton propagators in diagrams (a) and (b) go on-shell if the longitudinal momentum of the photon coupling to the spectator quarks vanishes (soft photon pole).
Second, particle propagators can go on-shell for a vanishing quark momentum (soft fermion pole).
This applies to one photon propagator in diagrams (a) and (b), and to the quark propagator in diagrams (c) and (d). 
It turns out, however, that the soft fermion pole contribution vanishes when summing over all diagrams.
Specifically, the contributions from diagrams (a) and (d) cancel each other out, and similarly for diagrams (b) and (c).
In this context we also refer to~\cite{Koike:2007dg} where such a cancellation has already been discussed.

The bottom part of the diagrams in Fig.$\,$\ref{f:twogamma_3parton} is expressed through a $q \gamma q$ correlator.
The discussion in the previous paragraph implies that we need this correlator only for the specific case of zero photon momentum.
Following the conventions of Ref.~\cite{Meissner:2009} we define the relevant soft photon pole matrix element $F_{FT}$ according to
\begin{equation}
\int \frac{d\xi^- d\zeta^-}{2(2\pi)^2} \, e^{ix P^+ \xi^-} 
\langle P,S | \bar{\psi}^{q}(0) \, \gamma^+ \, e F_{QED}^{+i}(\zeta) \, \psi^{q}(\xi)| P,S \rangle
= - M \varepsilon_T^{ij} S_T^{j} \, F_{FT}^q(x,x) \,,
\label{e:qpq_corr}
\end{equation}
with $\varepsilon_T^{ij} \equiv \varepsilon^{-+ij}$ $(\varepsilon_T^{12} = 1)$, and $e > 0$ denoting the elementary charge.
The photon is represented by a component of the electromagnetic field strength tensor.
The two arguments in $F_{FT}$ indicate the longitudinal quark momenta, which become equal for a vanishing photon momentum. 
Note that in Eq.~(\ref{e:qpq_corr}) Wilson lines between the field operators have been suppressed.

The $q \gamma q$ correlator $F_{FT}$ is the QED counterpart of the ETQS soft gluon pole matrix element $T_F$~\cite{Efremov:1981sh,Qiu:1991pp,Qiu:1998ia}.
In the notation of~\cite{Qiu:1998ia,Kouvaris:2006zy,Kang:2011hk}, $T_F$ is specified through
\begin{equation}
\int \frac{d\xi^- d\zeta^-}{4\pi} \, e^{ix P^+ \xi^-} 
\langle P,S | \bar{\psi}^{q}(0) \, \gamma^+ \, F_{QCD}^{+i}(\zeta) \, \psi^{q}(\xi)| P,S \rangle
= - \varepsilon_T^{ij} S_T^{j} \, T_{F}^q(x,x) \,.
\end{equation}
For later convenience we also recall the model-independent relation between $T_F$ and the transverse momentum dependent Sivers function $f_{1T}^{\perp}(x,\vec{p}_T^{\;2})$~\cite{Sivers:1989cc}. 
Taking the Sivers function as defined in semi-inclusive DIS (see, e.g., Ref.~\cite{Bacchetta:2006tn}) one has~\cite{Boer:2003cm,Ma:2003ut,Kang:2011hk}
\begin{equation}
- \, g \, T_{F}(x,x) = \int d^2 \vec{p}_T \, \frac{\vec{p}_T^{\;2}}{M} \, 
                  f_{1T}^{\perp}(x,\vec{p}_T^{\;2}) \Big|_{SIDIS} \,,
\label{e:ETQS_Sivers}                  
\end{equation}
with $g$ denoting the strong coupling constant.
If one instead considers the Sivers function appropriate for the Drell-Yan process the sign on the l.h.s.~in Eq.~(\ref{e:ETQS_Sivers}) has to be reversed~\cite{Collins:2002kn,Brodsky:2002rv}.
Furthermore, note that by definition the Sivers function and $g \, T_F$ do not depend on the sign of the strong coupling constant whereas $T_F$ does.

Let us now return to the target SSA $A_{UT}^N$.
The soft photon pole contribution from diagrams (a) and (b) in Fig.$\,$\ref{f:twogamma_3parton} gives rise to the following polarized cross section\footnote{We use here the same symbol for the polarized cross section as in Eq.~(\ref{e:pol_nucleon_1}) in order to avoid a proliferation of symbols.}:
\begin{eqnarray} 
k^{\prime 0} \frac{d\sigma_{pol}^{N}}{d^3\vec{k}'} & = & 
\frac{8 \pi \alpha_{em}^2 \, xy^2 \, M}{Q^8} \, \frac{\hat{s}^2 + \hat{t}^2}{\hat{u}^2} \,
\bigg( 2 + \frac{\hat{u}}{\hat{t}} \bigg) \,  
\varepsilon^{S_{N} P k k'} \, 
\sum_q e_q^2 \, x \tilde{F}_{FT}^{q/N}(x,x) \,,
\label{e:pol_nucleon_2} \\
&& \hspace{-2.4cm} \textrm{with} \quad \tilde{F}_{FT}(x,x) = F_{FT}(x,x) 
   - x \frac{d}{dx} F_{FT}(x,x) \,.
\nonumber
\end{eqnarray}
We performed the calculation both in the Feynman gauge and in the light-cone gauge in the collinear twist-3 approach.
We also compared the result to Ref.~\cite{Kouvaris:2006zy} where, in particular, the soft gluon pole contribution to the transverse SSA for the process $p^{\uparrow} p \to \pi X$ was studied.
Specifically, if one takes from that paper the result for the $q q' \to q' q $ channel, strips off the respective color factors, and pays attention to the sign conventions, one finds complete agreement between both calculations.
Note that for the QED treatment of the present work the overall sign of the polarized cross section is uniquely determined, since the sign of the photon-lepton coupling is fixed by means of the covariant derivative $D^{\mu} = \partial^{\mu} - i e A^{\mu}$. 
The so-called derivative term of the result, given by $(d/dx) F_{FT}$, and the non-derivative term, given by $F_{FT}$, have the same hard scattering coefficient, which is typical for such types of calculations~\cite{Kouvaris:2006zy,Koike:2007rq}.
If the momentum fraction $x$ is sufficiently large the derivative term numerically dominates over the non-derivative term~\cite{Qiu:1991pp,Qiu:1998ia}.
Like the unpolarized cross section in Eq.~(\ref{e:unp}), the expression in~(\ref{e:pol_nucleon_2}) contains an explicit factor of $\alpha_{em}^2$.
In addition, when evaluated, $F_{FT}$ is proportional to $\alpha_{em}$ so that the polarized cross section is actually proportional to $\alpha_{em}^3$ and hence suppressed.

On the basis of the polarized cross section in Eq.~(\ref{e:pol_nucleon_2}) and the unpolarized cross section in Eq.~(\ref{e:unp}), one can evaluate the SSA in Eq.~(\ref{e:ssa_def}).
We compute $A_{UT}^{N}$ in the target rest frame where one finds 
$\varepsilon^{S_{N} P k k'} = M \vec{S} \cdot (\vec{k} \times \vec{k}')$.
The lepton beam points along the negative $z$ direction $(\hat{k} = -\hat{e}_z)$, the $(xz)$ plane represents the lepton plane (where $\vec{k}'$ has a positive $x$ component), and the $y$ direction is defined according to $\hat{e}_y = \hat{e}_z \times \hat{e}_x$.
In that case the SSA is given by
\begin{equation}
A_{UT}^{N} = 
\frac{d\sigma(\uparrow_{y}) - d\sigma(\downarrow_{y})}{2 \, d\sigma_{unp}} =
- \frac{2\pi M}{Q} \, \frac{2-y}{\sqrt{1-y}} \, 
         \frac{\sum_q e_q^2 \, x \tilde{F}_{FT}^{q/N}(x,x)}
              {\sum_q e_q^2 \, x f_1^{q/N}(x)} \,,
\label{e:ssa_nucleon}
\end{equation}
with $\uparrow_{y} (\downarrow_{y})$ denoting polarization of the nucleon along $\hat{e}_y \, (- \hat{e}_y)$.
In Sec.~\ref{s:different_3} we will show numerical results for $A_{UT}^{N}$.
\subsection{Relation between {\boldmath $q \gamma q$} correlator and {\boldmath $qgq$} correlator}
\label{s:different_2}
\begin{figure}[t]
\begin{center}
\includegraphics[width=13.0cm]{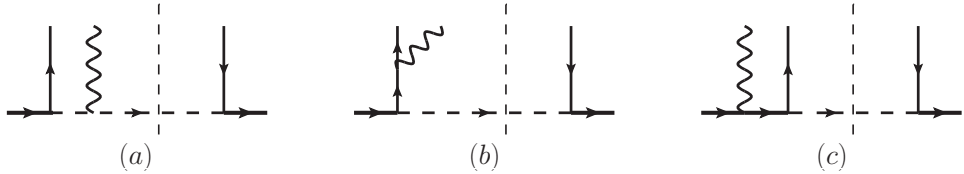} 
\end{center}
\caption{Feynman diagrams for the $q \gamma q$-correlation function $F_{FT}$ in a diquark model of the nucleon.
Diagram (c) has to be considered in the case of the proton.
Only diagram (a) gives a nonzero contribution.}
\label{f:diquark}
\end{figure}

We need some input for the $q \gamma q$ correlator $F_{FT}$ in order to obtain an estimate for $A_{UT}^{N}$.
To proceed we restrict ourselves to the large-$x$ valence quark region, i.e., we neglect in particular contributions from antiquarks.
For the comparison with the data from HERMES~\cite{:2009wj} and JLab~\cite{Katich:2011} this approximation should be justified.
Let us first assume that the two spectator quarks can be described through a single diquark.
In such a diquark model of the nucleon, $F_{FT}$ is given by the diagrams in Fig.$\,$\ref{f:diquark}.
It turns out that only diagram (a) gives a nonzero contribution.
By explicit calculation one can show that diagram (b) vanishes.
For a scalar diquark this was already pointed out in Ref.~\cite{Kang:2010hg}, while we extended this study to a vector diquark.
Moreover, also diagram (c) does not contribute to the soft photon pole matrix element as can readily be shown using contour integration.
These results hold for both a pointlike nucleon-quark-diquark vertex and a vertex containing an (arbitrary) form factor.

Since we are left with diagram (a) only, one can establish a simple relation between $F_{FT}$ and the ETQS matrix element $T_F$ for which the same graph has to be computed with the photon replaced by a gluon.
The precise form of the relation for a given quark and nucleon mainly depends on the strong coupling and the electromagnetic charge of the diquark. 
One obtains
\begin{eqnarray}
&& F_{FT}^{u/p} = - \, \frac{\alpha_{em}}{6 \pi C_F \alpha_s M} \, (g \, T_F^{u/p}) \,,
\qquad \qquad
F_{FT}^{d/p}  =  - \, \frac{2 \, \alpha_{em}}{3 \pi C_F \alpha_s M} \, (g \, T_F^{d/p}) \,,
\nonumber \\
&& F_{FT}^{u/n} =   \frac{\alpha_{em}}{3 \pi C_F \alpha_s M} \, (g \, T_F^{d/p}) \,,
\hspace{1.85cm}
F_{FT}^{d/n}  = - \, \frac{\alpha_{em}}{6 \pi C_F \alpha_s M} \, (g \, T_F^{u/p}) \,,
\label{e:relations}
\end{eqnarray}
where $C_F = 4/3$.
Note that for $T_F^{q/N}$ we used isospin symmetry.
We assumed a pointlike coupling between the gauge bosons and the diquark.
Since the gauge boson has a vanishing (longitudinal) momentum this assumption appears reasonable to us. 
We point out that exactly the same relations between $F_{FT}$ and $T_F$ are valid in a more general 3-quark picture of the nucleon.
To arrive at this result one merely needs the fact that it does not matter to which of the two spectator quarks the gauge boson couples.
This obviously holds if the spectator quarks have the same flavor, but it also applies in the case of different flavors~\cite{Pasquini:2010af}\footnote{We also acknowledge a discussion with C.~Lorc\'e and B.~Pasquini about that point.}.
If higher order corrections are taken into consideration the simple relations in~(\ref{e:relations}) break down.
However, we do not expect such corrections to affect the general conclusions drawn in Sec.~\ref{s:different_3}.
We also mention that, when computing the target SSAs, we evaluate the strong coupling constant appearing in~(\ref{e:relations}) at the scale $\mu^2 = Q^2$.

Before discussing the numerical results we briefly explain why the diagrams with the photons coupling to different quarks presumably dominate over those considered in Sec.~\ref{s:same}.
The total contribution to the polarized cross section, caused by the diagrams where the photons couple to the same quark, is proportional to $\alpha_{em}^3$ multiplied by $qgq$ correlators like $g T_F$~\cite{Schlegel:prog}. 
(Here we neglect the transversity-related quark mass term.) 
This, however, is suppressed by a factor $\alpha_s$ compared to the diagrams in Fig.$\,$\ref{f:twogamma_3parton} --- see Eqs.~(\ref{e:pol_nucleon_2}) and (\ref{e:relations}).
Though one might argue that this factor $\alpha_s$ should be evaluated in the non-perturbative regime it may well give rise to a numerical suppression.
Second, it has been shown that for elastic lepton-nucleon scattering photons coupling to different quarks even dominate in a $1/Q$ expansion over those coupling to the same quark~\cite{Borisyuk:2008db,Kivel:2009eg}.
(This finding is equivalent to the so-called Landshoff mechanism for elastic scattering of hadrons at large momentum transfer~\cite{Landshoff:1974ew}.)
Therefore, one may expect a similar behavior for inclusive DIS in the region of larger values of $x$.
Of course, only a full calculation of the diagrams containing $qgq$ correlations~\cite{Schlegel:prog} will provide an ultimate answer on this point.
\subsection{Numerical results and discussion}
\label{s:different_3}
For our numerical estimates we have used three inputs for the $q \gamma q$ correlator $F_{FT}$ based on three different extractions of $T_F$:
\begin{itemize}
\item {\bf Sivers} input: This input uses an indirect extraction of $T_F$ based on the Sivers functions obtained from HERMES~\cite{Diefenthaler:2007rj} and COMPASS~\cite{Martin:2007au} data on production of pions and kaons in semi-inclusive DIS~\cite{Anselmino:2008sga}, and the relation in Eq.~(\ref{e:ETQS_Sivers}) between $f_{1T}^{\perp}$ and $T_F$.
\item {\bf KQVY} input: This input uses a direct extraction of $T_F$ by Kouvaris, Qiu, Vogelsang, and Yuan (KQVY)~\cite{Kouvaris:2006zy}, based on Fermilab data ($\sqrt{s} \approx 20 \, \textrm{GeV}$) for $p^{\uparrow} p \to \pi X$ and $\bar{p}^{\uparrow} p \to \pi X$~\cite{Adams:1991rw}, and RHIC data ($\sqrt{s} = 200 \, \textrm{GeV}$) from STAR~\cite{Adams:2003fx,Abelev:2008af} (for $p^{\uparrow} p \to \pi X$) and BRAHMS~\cite{Arsene:2008aa} (for $p^{\uparrow} p \to \pi X$ and $p^{\uparrow} p \to K X$). 
(See also Ref.~\cite{Adamczyk:2012xd} for the most recent STAR data on $p^{\uparrow} p \to \pi X$ and $p^{\uparrow} p \to \eta X$.)  
We took into account the sign error of that extraction~\cite{Kang:2011hk}.
Also, we used Fit I in Ref.~\cite{Kouvaris:2006zy} which contains valence quarks only.
Our general conclusions would not change if we used Fit II from~\cite{Kouvaris:2006zy} where sea quarks are considered as well.
It has been pointed out that the Sivers and the KQVY extractions of $T_F$ disagree in sign~\cite{Kang:2011hk}.
Currently, this sign mismatch issue represents a key puzzle in hadronic spin physics.
Below we will come back to that puzzle.
\item {\bf KP} input: This input also uses an indirect extraction of $T_F$ based on a new fit of the Sivers function by Kang and Prokudin (KP)~\cite{Kang:2012xf}, and the relation in Eq.~(\ref{e:ETQS_Sivers}).
The authors of~\cite{Kang:2012xf} attempted a simultaneous fit of recent HERMES~\cite{Airapetian:2009ae} and COMPASS~\cite{Alekseev:2008aa} data on pion production in semi-inclusive DIS, as well as RHIC data on $p^{\uparrow} p \to \pi X$~\cite{Abelev:2008af,Arsene:2008aa}.
They allowed for the Sivers function to have a node in $x$ and in $p_T$ --- see also the recent work~\cite{Boer:2011fx,Bacchetta:2011gx} where a node of $f_{1T}^{\perp}$ in $x$ was discussed.
Even when considering such nodes it was found that, with the Sivers effect alone, no satisfactory combined description of SSAs in semi-inclusive DIS and in proton-proton collisions can be achieved. 
In other words, Sivers functions with nodes do not help to resolve the aforementioned sign mismatch problem.
\end{itemize}
\begin{figure}[t]
\begin{center}
\includegraphics[width=7.5cm]{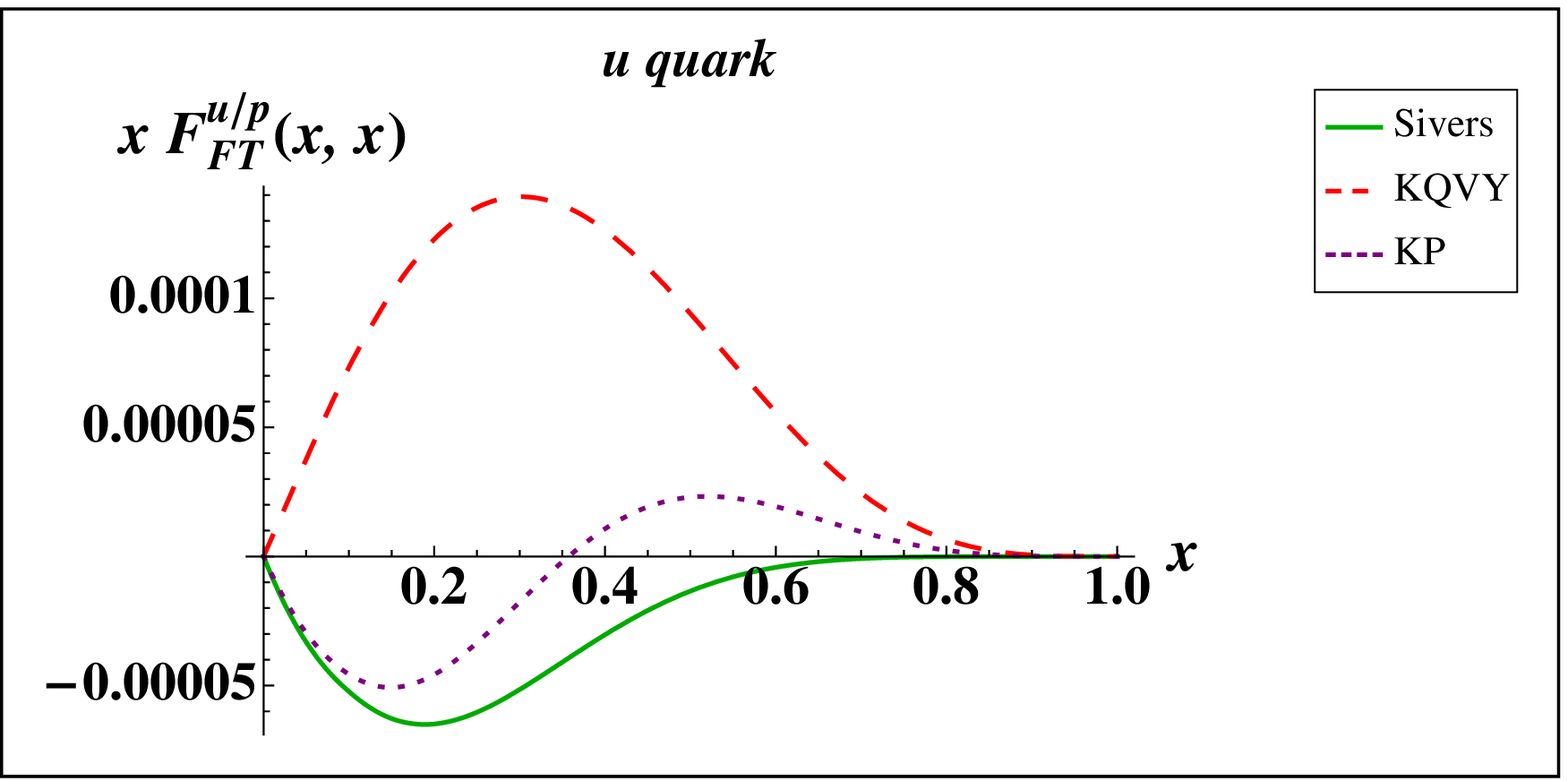} 
\hspace{1.0cm} 
\includegraphics[width=7.5cm]{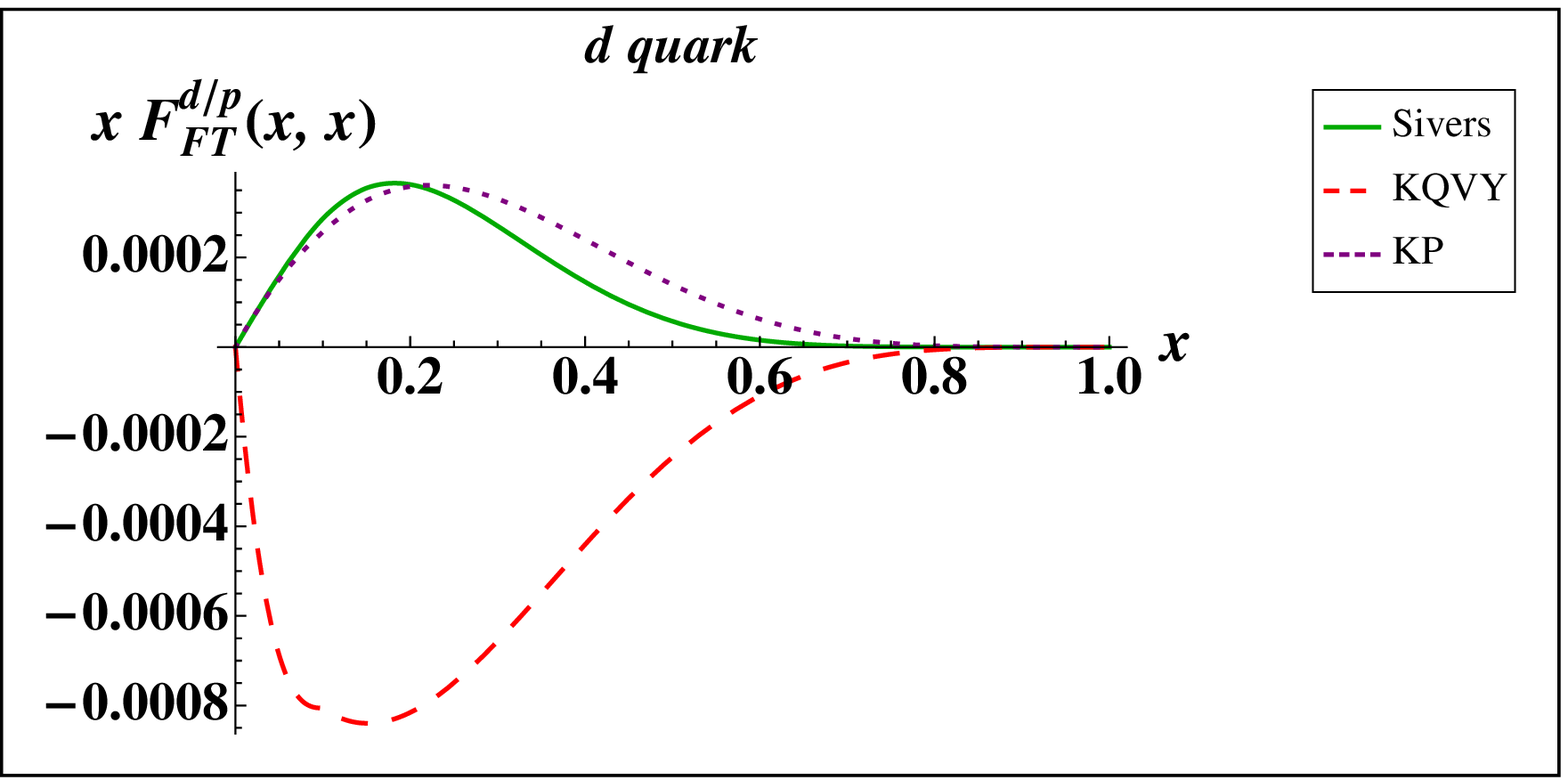} 
\end{center}
\caption{Numerical result for our model for the $q \gamma q$ correlator $F_{FT}$ for the proton at the scale $\mu^2 = 2\,\textrm{GeV}^2$, based on three different inputs for $T_F$.
Left panel: Up quarks. Right panel: Down quarks.}
\label{f:FFT}
\end{figure}

The numerical results for $F_{FT}$ for the proton are shown in Fig.$\,$\ref{f:FFT}.
The neutron results can be obtained immediately from these plots too since, according to~(\ref{e:relations}), one has $F_{FT}^{u/n} = - F_{FT}^{d/p} / 2$ and $F_{FT}^{d/n} = F_{FT}^{u/p}$.
In general, the magnitude of $F_{FT}$ is rather small, which is mainly due to the factor $\alpha_{em}$ showing up in~(\ref{e:relations}). 
The opposite sign for the Sivers input and the KQVY input nicely reflects the sign mismatch holding for both up quarks and down quarks~\cite{Kang:2011hk}.
Also, notice the node in $F_{FT}^{u/p}$ for the KP input which is a direct consequence of a node in the up quark Sivers function~\cite{Kang:2012xf} --- see Eqs.~(\ref{e:relations}) and~(\ref{e:ETQS_Sivers}).
Because of this node, the KP fits for $f_{1T}^{\perp u/p}$ and $f_{1T}^{\perp d/p}$ have the same sign at larger values of $x$~\cite{Kang:2012xf}.
Such a  scenario is actually at variance with a model-independent large-$N_c$ analysis~\cite{Pobylitsa:2003ty}.

\begin{figure}[t]
\begin{center}
\includegraphics[width=7.5cm]{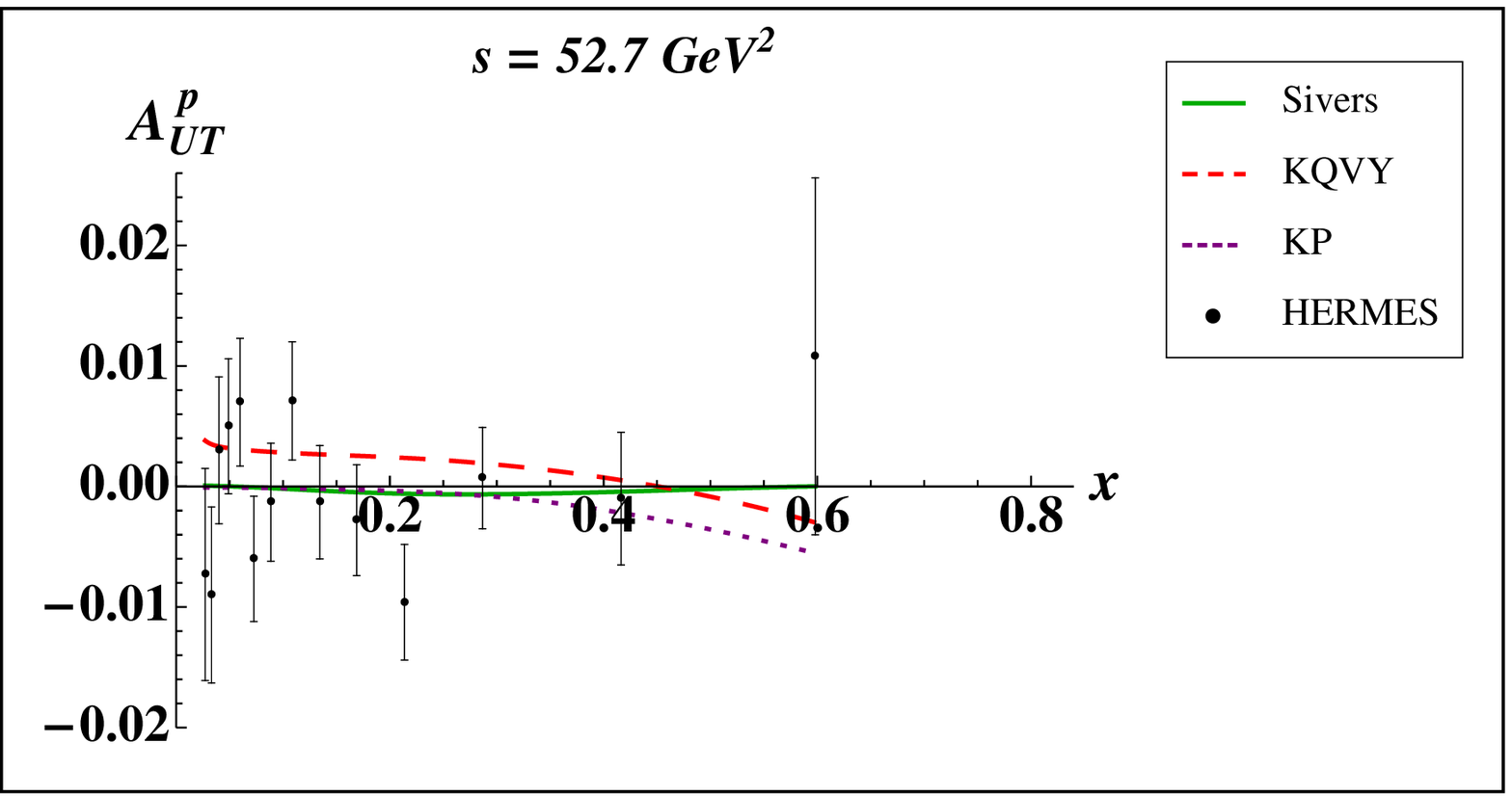} 
\hspace{1.0cm} 
\includegraphics[width=7.5cm]{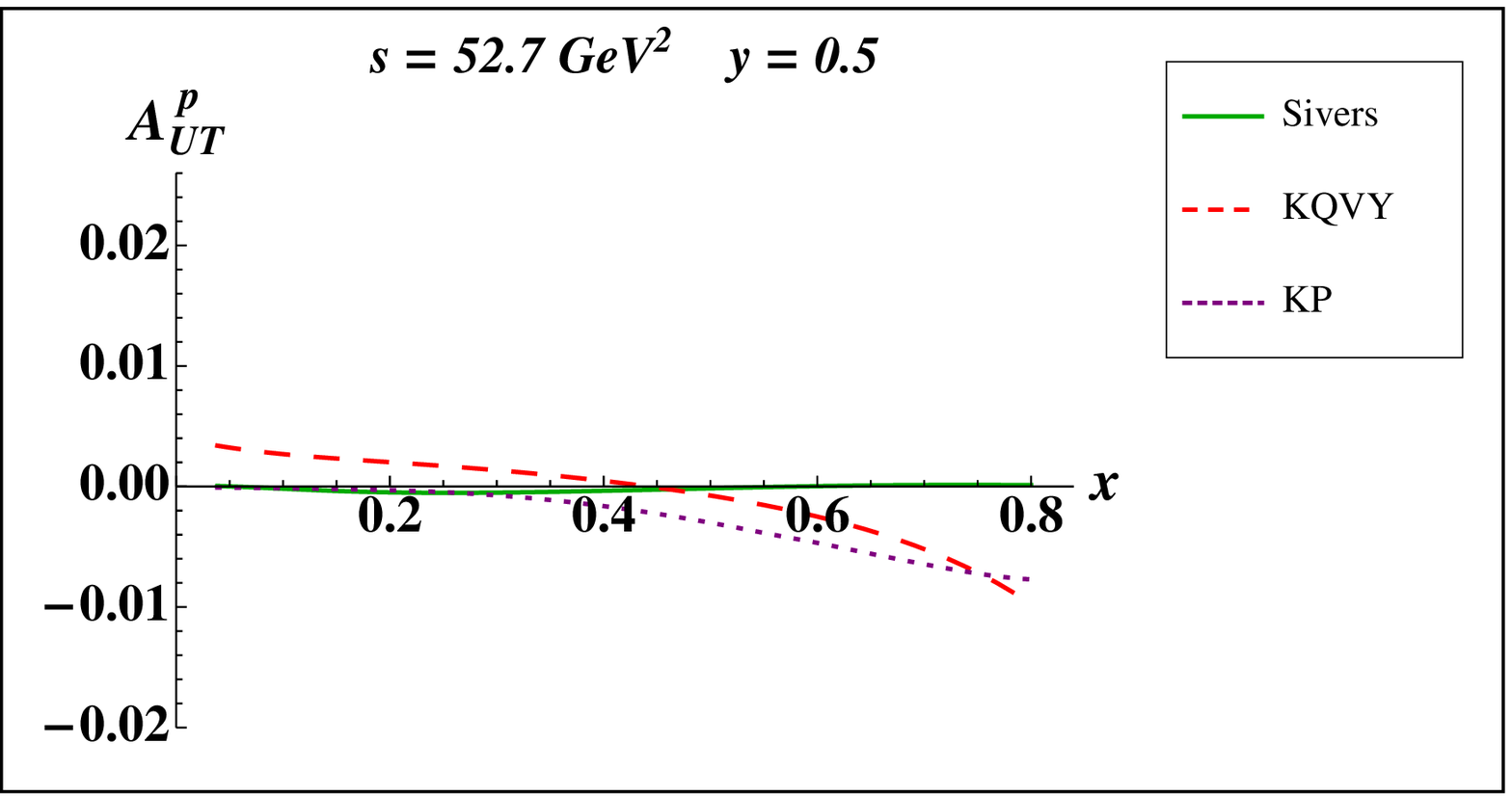} 
\end{center}
\caption{Proton target asymmetry for three inputs for $F_{FT}$.
Left panel: Comparison to data from the HERMES Collaboration~\cite{:2009wj}.
The asymmetry has been evaluated for the average values $\langle x \rangle$ and $\langle Q^2 \rangle $ of the data bins.
Right panel: Results for typical HERMES kinematics at $y = 0.5$, where the lowest $x$ value for the curves corresponds to $Q^2 = 1\,\textrm{GeV}^2$.}
\label{f:AUT_pro}
\end{figure}

Let us now turn our attention to the target asymmetry in Eq.~(\ref{e:ssa_nucleon}).
When computing the asymmetries, for the unpolarized parton densities $f_1^q$ we take the GRV98 parametrization~\cite{Gluck:1998xa} for the Sivers input and the KP input, whereas we take the CTEQ5L parametrization~\cite{Lai:1999wy} in the case of the KQVY input.
Those parametrizations were the ones used in the respective fits.
The results for $A_{UT}^{p}$ are displayed in Fig.$\,$\ref{f:AUT_pro}.
From the plot on the left panel of that figure we see that the Sivers input perfectly agrees with the HERMES data.
Overall, also the KQVY and the KP inputs give a good description of the data.
One may, however, sense from the plot that these two inputs become somewhat too large towards larger $x$.
In the KP case this is due to the aforementioned node in $f_{1T}^{\perp u/p}$.
Therefore, our results indicate that such a node is not preferred, even though it is not ruled out by the current data either.
The plot on the right panel of Fig.$\,$\ref{f:AUT_pro} shows results for typical HERMES kinematics extending until $x = 0.8$.
This figure suggests that sufficiently precise data for $A_{UT}^{p}$ at such large $x$ values may allow one to distinguish between the different inputs for $F_{FT}$ and, hence, the different inputs for $T_F$.
(Of course, here one has to keep in mind that also the Sivers input is hardly constrained for $x > 0.4$~\cite{Anselmino:2008sga, Diefenthaler:2007rj, Martin:2007au}.)
We note that the KP input actually provides a finite asymmetry in the limit $x \to 1$, while for the KQVY input the asymmetry diverges in that limit.
In this context we would like to add the following point:
in general, if one fits available data on SSAs in processes like $p^{\uparrow} p \to \pi X$ by means of the Sivers effect (as described in the twist-3 collinear framework) alone, then the resulting $T_F$ typically is such that the asymmetry violates the positivity bound for $x_F \to 1$~\cite{Kouvaris:2006zy,Kanazawa:2010au}.
(Note that if the so extracted $T_F$ is used to compute SSAs for other processes, the positivity bound may be violated already at lower values of $x_F$ --- in this context, see for instance~\cite{Koike:2002ti}.) 
This phenomenon, which is caused by the strong rise of the measured asymmetries towards larger values of $x_F$, gives some hint that the twist-3 Sivers-type effect may not be the only cause of the observed SSAs.
As we discuss in more detail below, the data on the neutron SSA in inclusive DIS clearly support this point of view.

\begin{figure}[!]
\begin{center}
\includegraphics[width=7.5cm]{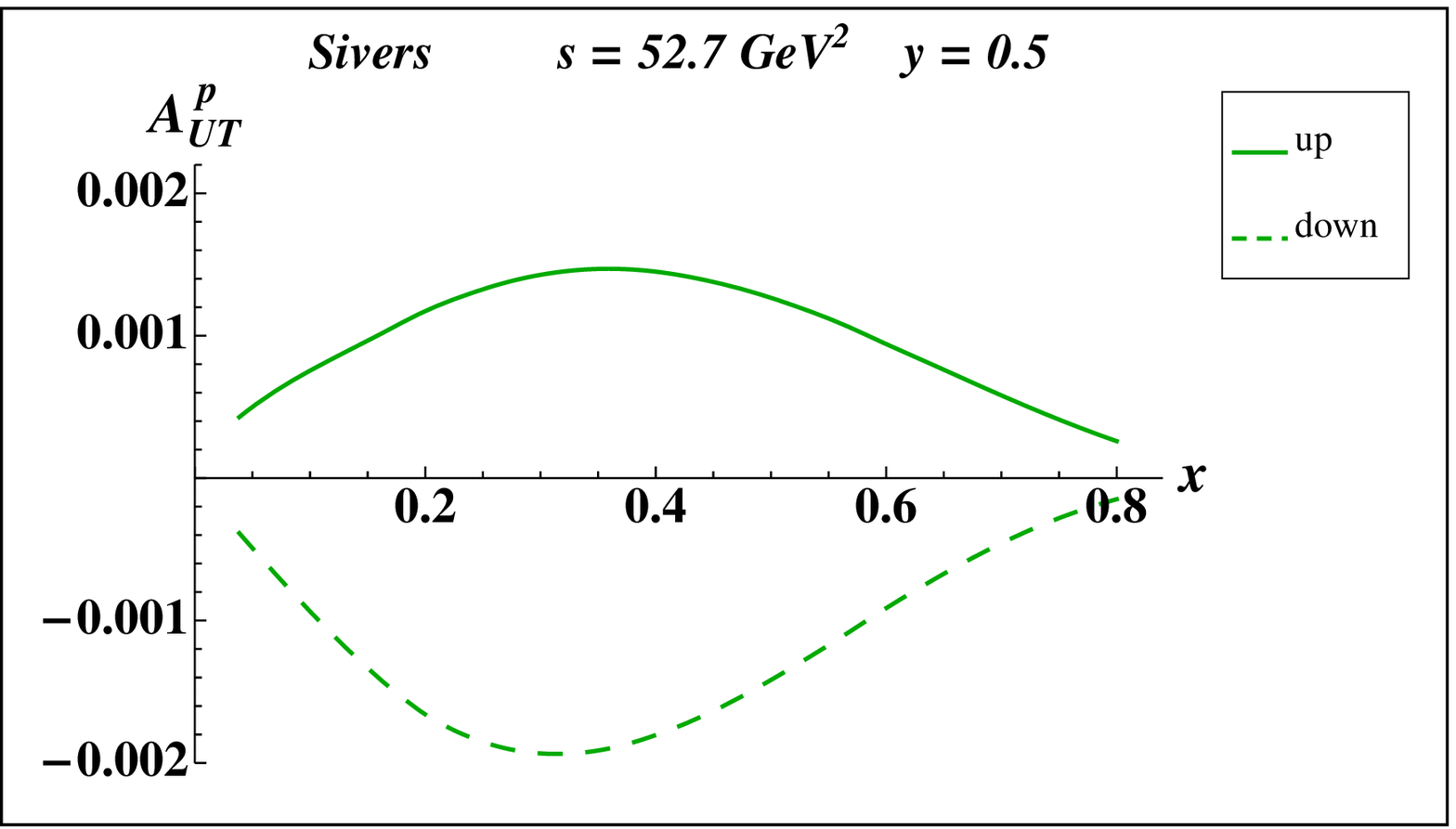} 
\hspace{1.0cm} 
\includegraphics[width=7.5cm]{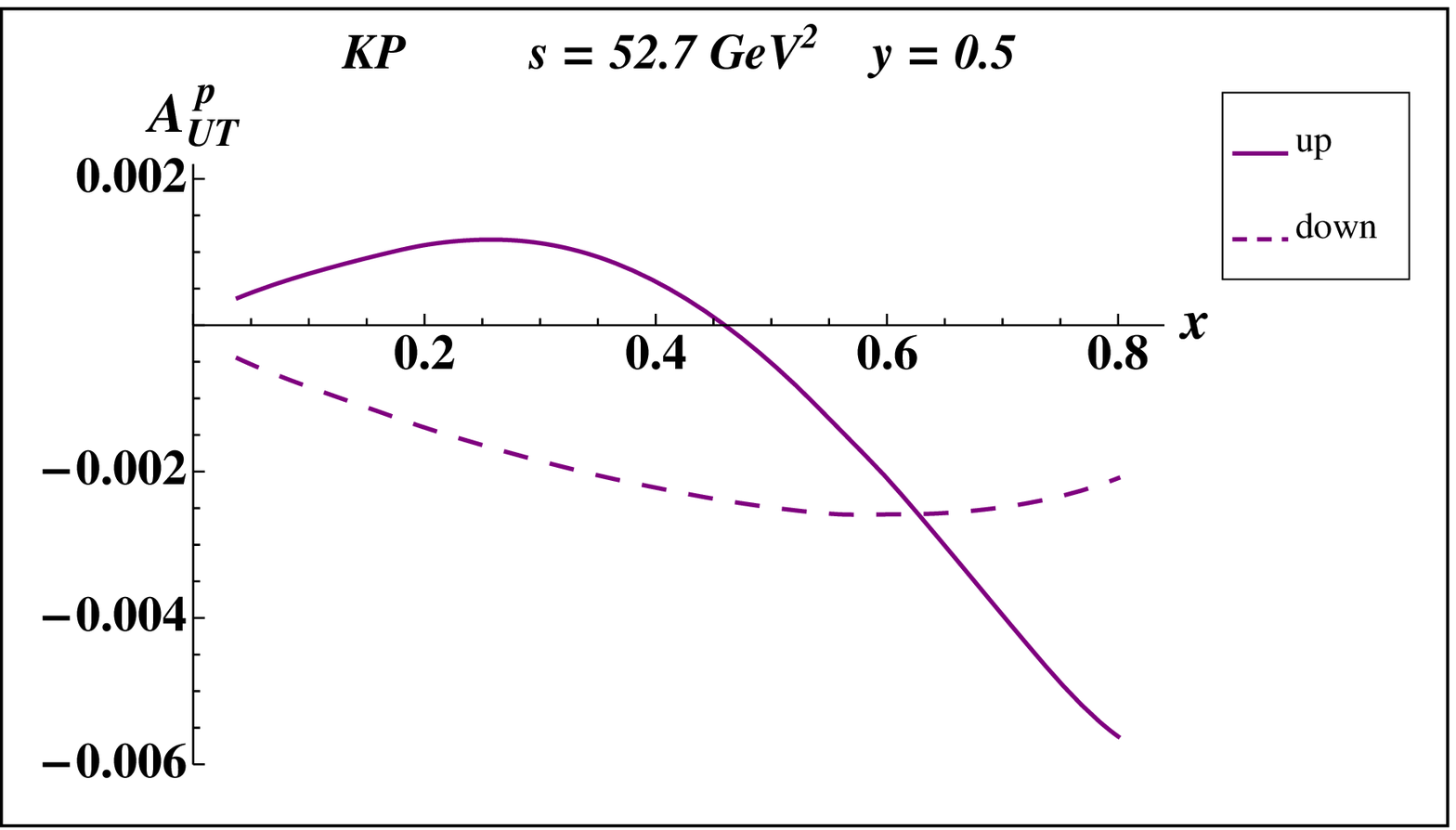} 
\end{center}
\caption{Individual flavor contributions to the proton asymmetry for the Sivers input (left panel) and the KP input (right panel).
The results are for typical HERMES kinematics at $y = 0.5$, where the lowest $x$ value for the curves corresponds to $Q^2 = 1\,\textrm{GeV}^2$.}
\label{f:pro_flavor}
\end{figure}

Before moving on to the neutron case it is instructive to look at the individual flavor contributions to $A_{UT}^{p}$ which are shown for the Sivers and KP inputs in Fig.$\,$\ref{f:pro_flavor}. 
In the Sivers case these individual contributions are quite small and, moreover, they almost exactly cancel each other.
This explains the tiny effect we find for $A_{UT}^{p}$.
(Note that other available extractions of the Sivers functions from data on semi-inclusive DIS~\cite{Efremov:2004tp, Anselmino:2005nn, Vogelsang:2005cs, Collins:2005ie, Arnold:2008ap, Bacchetta:2011gx,Anselmino:2012aa} lead to a similarly small effect for $A_{UT}^{p}$.)
For the KP input there is also a partial cancellation among the flavor contributions at lower values of $x$, whereas at larger $x$, because of the node in $x$ in $f_{1T}^{\perp u/p}$, up quark and down quark contributions give rise to the same sign.

\begin{figure}[t]
\begin{center}
\includegraphics[width=7.5cm]{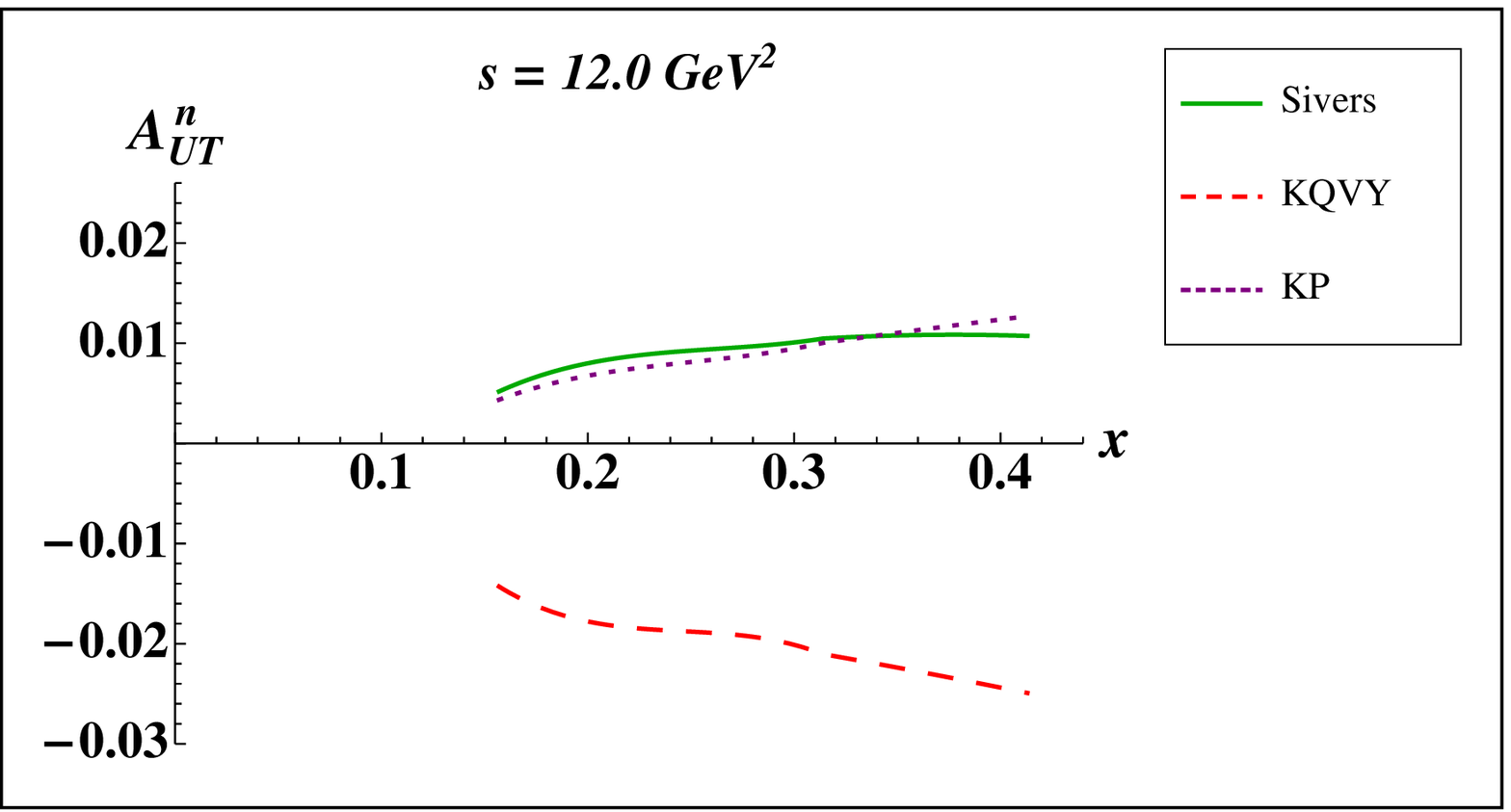} 
\hspace{1.0cm} 
\includegraphics[width=7.5cm]{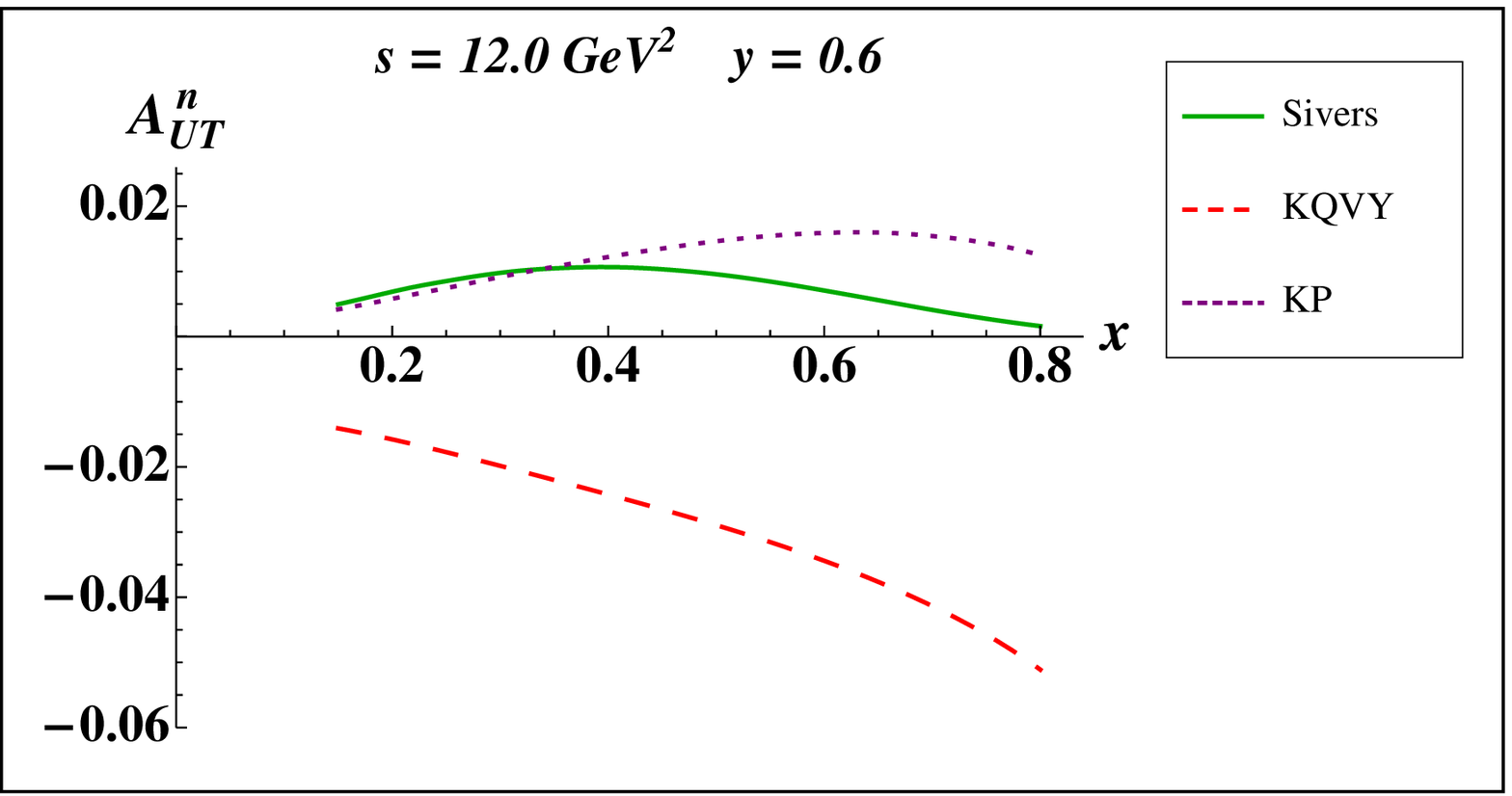} 
\end{center}
\caption{Neutron target asymmetry for three inputs for $F_{FT}$.
Left panel: Calculation for kinematics of the experiment in Hall A at Jefferson Lab~\cite{E07-013,Katich:2011}.
The asymmetry has been evaluated for the average values $\langle x \rangle$ and $\langle Q^2 \rangle $ of the data bins.
The preliminary result for $A_{UT}^n$ is positive and on the (few) percent level~\cite{Katich:2011}.
Right panel: Results for typical Jefferson Lab kinematics at $y = 0.6$, where the lowest $x$ value for the curves corresponds to $Q^2 = 1\,\textrm{GeV}^2$.}
\label{f:AUT_neu}
\end{figure}

\begin{figure}[!]
\begin{center}
\includegraphics[width=7.5cm]{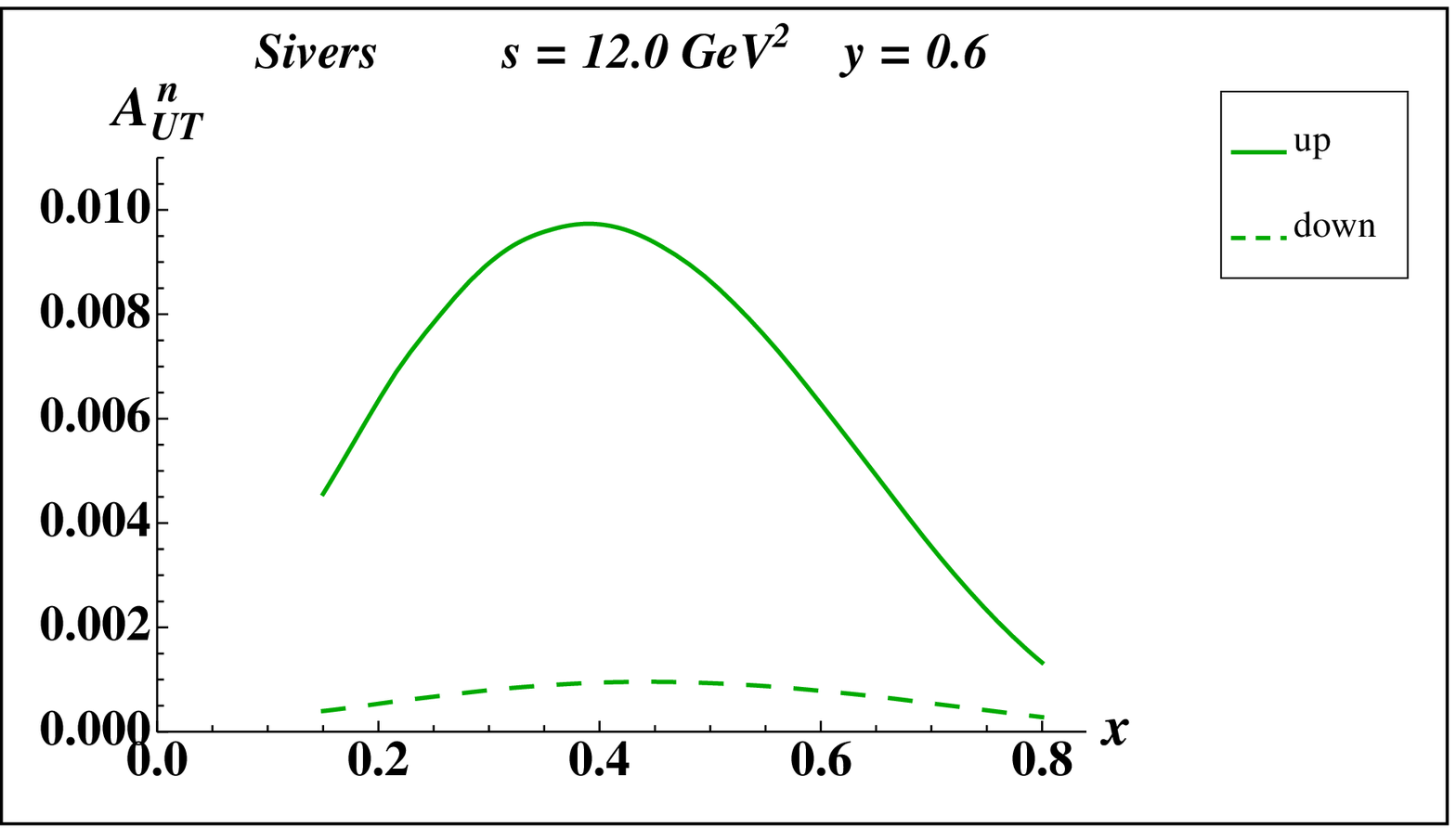} 
\hspace{1.0cm} 
\includegraphics[width=7.5cm]{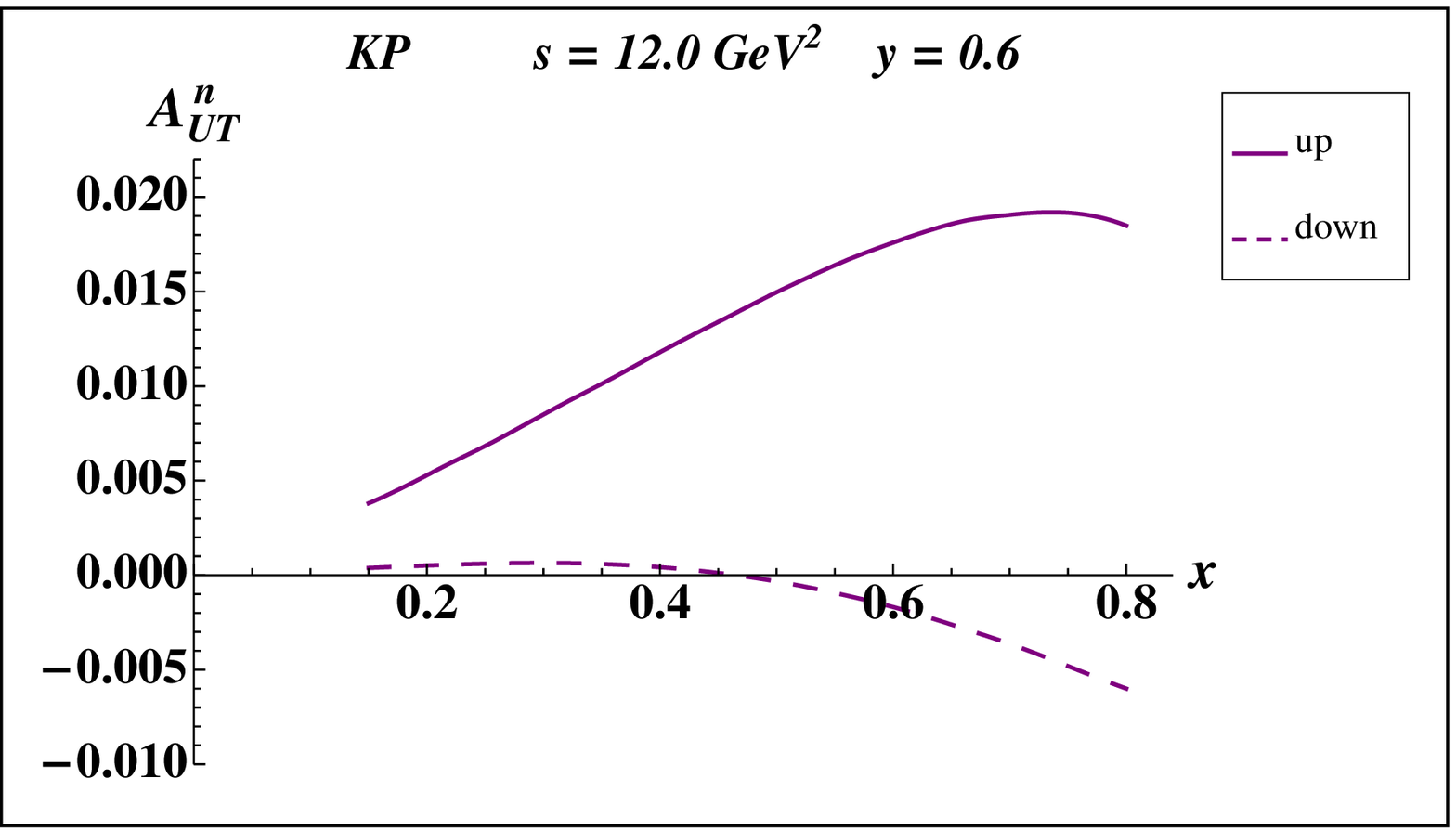} 
\end{center}
\caption{Individual flavor contributions to the neutron asymmetry for the Sivers input (left panel) and the KP input (right panel).
The results are for typical Jefferson Lab kinematics at $y = 0.6$, where the lowest $x$ value for the curves corresponds to $Q^2 = 1\,\textrm{GeV}^2$.}
\label{f:neu_flavor}
\end{figure}

In Fig.$\,$\ref{f:AUT_neu}, the results for the neutron asymmetry $A_{UT}^{n}$ are shown.
While the plot on the left panel of that figure is for the kinematics of the E07-013 experiment in Hall A at Jefferson Lab~\cite{E07-013,Katich:2011}, the one on the right panel shows results for typical Jefferson Lab kinematics at fixed values of $y$ covering a larger $x$ range.
We refrain from including in the plot the data given in Ref.~\cite{Katich:2011}, since final data from the Hall A Collaboration should be available soon~\cite{Private:ACK:2012}.
The (preliminary) data from~\cite{Katich:2011} on $A_{UT}^{n}$ are on the (few) percent level and positive. 
In fact the Sivers and KP inputs agree reasonably well with these data.
(In the $x$ region of the current data both inputs actually provide almost identical results and differences become manifest only towards larger $x$ --- see plot on the right panel of Fig.$\,$\ref{f:AUT_neu}.) 
This means, in particular, that we have a framework which can simultaneously describe both the vanishing/tiny result for $A_{UT}^{p}$ as well as the nonzero result for $A_{UT}^{n}$. 
The plots of the individual flavor contributions in Fig.$\,$\ref{f:neu_flavor} show that the neutron asymmetry is dominated by the (relatively large) up quark contribution.

Notice that if we had a node in $p_T$ in the Sivers function such that $f_{1T}^{\perp}$ (at low $p_T$) and the particular $p_T$ moment of the Sivers function on the r.h.s.~in Eq.~(\ref{e:ETQS_Sivers}) have opposite signs, we would obtain the wrong sign for the neutron asymmetry from the Sivers input.
Such a node was suggested in Ref.~\cite{Kang:2011hk} as a potential resolution of the sign mismatch problem.
However, as a result of a more detailed phenomenological work, it was already pointed out in Ref.~\cite{Kang:2012xf} that a node of the Sivers function in $p_T$ can hardly solve the sign mismatch issue.  
Our finding here is in line with that observation. 

An additional crucial lesson from the neutron data is that the KQVY input gives us the wrong sign for $A_{UT}^{n}$.
This suggests that the SSAs seen in processes like $p^{\uparrow} p \to \pi X$ are not mainly caused by the Sivers mechanism as described in twist-3 collinear factorization.
Therefore, the sign mismatch problem boils down to the question about the origin of the SSAs in those reactions.
One popular alternative mechanism considered in the literature is the Collins effect~\cite{Collins:1992kk} (or its collinear twist-3 analog), which describes a transverse SSA in parton fragmentation.
For instance, in a recent study carried out in a generalized (transverse momentum dependent) parton model it was argued that the Collins effect can contribute significantly but is not sufficient to entirely describe the observed effects~\cite{Anselmino:2012rq}.
This is an important result even though the existing papers on this topic --- see~\cite{Kang:2010zzb,Kang:2011ni,Anselmino:2012rq} and references therein --- do not allow one to draw an ultimate conclusion about the role of the fragmentation-related contribution to SSAs in hadronic collisions.
Other mechanisms have been proposed as well~\cite{Hoyer:2006hu,Hoyer:2008fp,Qian:2011ya,Kovchegov:2012ga}, and more work is required in order to settle this important question.

Let us also remark that at least a qualitative simultaneous description of SSAs in semi-inclusive DIS and in hadronic collisions can be achieved if one merely assumes the existence of a Sivers function without relating $f_{1T}^{\perp}$ to the rescattering of active partons with beam remnants~\cite{Boglione:2007dm,Anselmino:2009hk}.
In such a scenario, in particular, one does not face a sign mismatch problem between the different processes.
However, if the rescattering picture, which also underlies the current definition of transverse momentum dependent parton distributions in QCD~\cite{Brodsky:2002cx,Collins:2002kn}, is taken seriously, then the sign mismatch problem cannot be circumvented~\cite{Kouvaris:2006zy,Gamberg:2010tj,Kang:2011hk,Kanazawa:2011bg}. 
The fact that in the present work we can describe SSAs in inclusive DIS using the Sivers function extracted from semi-inclusive DIS does support the rescattering picture underlying the Sivers effect --- see diagrams (a) and (b) in Fig.~\ref{f:twogamma_3parton}.
Obviously, our analysis also supports that the observed so-called $\sin(\phi_h - \phi_S)$ modulation of the cross section for semi-inclusive DIS can indeed be associated with the Sivers function.

Finally, we briefly comment on two potential error sources of our numerical results in addition to those caused, e.g., by the model for the $q \gamma q$ correlator $F_{FT}$.
First, the three different extractions of $T_F$ that we are using have uncertainties.
However, only for the Sivers input is a quantitative error estimate available~\cite{Anselmino:2008sga}.
Since the neutron asymmetry is almost entirely determined by the down quark Sivers function for the proton --- see plot on the left panel of Fig.~\ref{f:neu_flavor} as well as Eq.~(\ref{e:relations}) --- the error for $A_{UT}^{n}$ in the case of the Sivers input corresponds to the error for $f_{1T}^{\perp d/p}$.
According to Ref.~\cite{Anselmino:2008sga} the error for $f_{1T}^{\perp d/p}$, in the $x$ range of the Jefferson Lab data, is roughly of the order $\pm(30 - 50) \%$ with the larger uncertainty towards larger values of $x$.
This suggests that the Sivers input leads to a robust prediction of a positive $A_{UT}^{n}$. 
Second, target mass corrections may play a role, especially for the Jefferson Lab kinematics.
While it is definitely worth trying to obtain an estimate of such corrections, a corresponding study would go beyond the scope of this work, not the least because at present there exists no generally accepted framework for treating target mass corrections in DIS --- see Refs.~\cite{Accardi:2008ne,Accardi:2008pc,Accardi:2009md,Brady:2011uy,Steffens:2012jx} for detailed recent work on this topic.
Moreover, since the polarized cross section entering $A_{UT}$ is a twist-3 observable, it is not clear if any of the proposed frameworks in the literature would be directly applicable.
Target mass corrections might also cancel to a large extent in the spin asymmetry.

%
%
\section{Summary}
\label{s:summary}

In summary, we have investigated transverse SSAs in inclusive DIS off the nucleon with a focus on the target SSA, which recently has been measured for a proton target~\cite{:2009wj} and a neutron target~\cite{E07-013,Katich:2011}.
Such observables can exist only if more than one photon is exchanged between the leptonic and the hadronic part of the process.
In the present work we have considered two-photon exchange where the two photons couple to {\it different} quarks inside the nucleon, and we have given arguments why this mechanism presumably dominates over two photons coupling to the {\it same} quark considered in Refs.~\cite{Metz:2006pe,Afanasev:2007ii,Schlegel:prog}.

Our calculation has been carried out in the collinear twist-3 approach and contains a quark-photon-quark correlator (denoted by $F_{FT}$) of the nucleon. 
We have shown that in a valence quark picture of the nucleon $F_{FT}$ is related to the Efremov-Teryaev-Qiu-Sterman quark-gluon-quark correlator $T_F$.
The function $T_F$ has a model-independent relation to the transverse momentum dependent Sivers function $f_{1T}^{\perp}$ and plays a crucial role in the QCD description of SSAs for processes like $p^{\uparrow} p \to \pi X$.

The relation between $F_{FT}$ and $T_F$ has allowed us to numerically compute the target SSA in inclusive DIS.
We have found a reasonable description of both the proton data and the (preliminary) neutron data, provided that we take $T_F$ obtained from a fit of the Sivers function to data in semi-inclusive DIS~\cite{Anselmino:2008sga}.
Our analysis also indicates that a node of $f_{1T}^{\perp}$ in $x$ is not preferred, though it is not ruled out either by the currently available data.
We also argued that a node of the Sivers function in the transverse momentum $p_T$ such that $f_{1T}^{\perp}$ (at low $p_T$) and the $p_T$ moment of $f_{1T}^{\perp}$ on the r.h.s.~in Eq.~(\ref{e:ETQS_Sivers}) have opposite signs, would not work --- see also Ref.~\cite{Kang:2012xf}.
Moreover, our study suggests that the nonzero $\sin(\phi_h - \phi_S)$ modulation observed in semi-inclusive DIS can be attributed to the Sivers function and, in particular, supports the understanding that the Sivers effect is intimately related to the reinteraction of active partons with target remnants.
If we take $T_F$ based on a direct extraction from SSAs in hadronic collisions~\cite{Kouvaris:2006zy,Kang:2011hk}, we obtain the wrong sign for the neutron asymmetry in inclusive DIS.
This finding indicates that the Sivers effect cannot be the only source of the SSAs observed in processes like $p^{\uparrow} p \to \pi X$.
Keeping in mind the large spin effects (up to $50\,\%$) measured there, it is of crucial importance to eventually settle what causes these asymmetries.
\\[0.5cm]
%
%
\noindent
{\bf Acknowledgments:}
We are grateful to J.~Katich for sending us his Ph.D. thesis.
Moreover,  we thank D.~Hasch, A.~L\'opez Ruiz, A.~Mart\'inez de la Ossa, and G.~Schnell for a discussion about the sign conventions of the HERMES data, and J.-P.~Chen for a discussion about the data from Jefferson Lab.
This work has been supported in part by the NSF under Grant No.~PHY-0855501, and by the BMBF under Grant No.~OR 06RY9191.


\end{document}